University of South Florida, Sarasota- Manatee

College of Hospitality and Tourism leadership

# A STUDY OF CYBER SECURITY IN HOSPITALITY INDUSTRY ─ THREATS AND COUNTERMEASURES: CASE STUDY IN RENO, NEVADA

Master of Science in Hospitality Management

By Neda Shabani

Project Supervisor: Dr. Cihan Cobanoglu

April 2016

*ABSTRACT.* The purpose of this study is to analyze cyber security and security practices of electronic information and network system, network threats, and techniques to prevent the cyber attacks in hotels. Helping the information technology directors and chief information officers (CIO) is the aim of this study to advance policy for security of electronic information in hotels and suggesting some techniques and tools to secure the computer networks. This research is completely qualitative while the case study and interviews have done in 5 random hotels in Reno, Nevada, United States of America. The interview has done with 50 hotel guests, 10 front desk employees, 3 IT manager and 2 assistant of General manager. The results show that hotels' cyber security is very low and hotels are very vulnerable in this regard and at the end, the implications and contribution of the study is mentioned.

## *INTRODUCTION*

"Modern societies have become increasingly dependent on Information and Communication Technologies that offer both opportunities and challenges with respect to improvements in the quality of life of people and the communities in which they live" (Muata, Bryson, Vogel, 2014, p.93). Technology in hospitality industry is driven by the increasing transaction volumes, complex reporting requirement, and international communication needs (Cobanoglu, & Demicco, 2007). Information technology (IT) can improve almost all areas of hospitality industry, such as guest services, reservations, food and beverage management, sales, food service catering, maintenance, security, and hospitality accounting (Cobanoglu, & Demicco, 2007).

Having high technology causes a lot of data breach and information loss. To prevent against losses, organizations observe their computer network for security threats from a wide range of sources, such as computer-assisted fraud, espionage, sabotage, vandalism, hacking, system failures, fire, and flood (Cobanoglu, & Demicco, 2007). To protect the public trust and to prevent copycat hackers to hack into an organization's computer system, most of the organizations try not to reveal the data breaches and cyber attacks against their computer systems (Cobanoglu, & Demicco, 2007).

*Problem Statement*

Every level of hotel management (e.g., property level, user level) involves computer networks. At the property level, there are local area networks where reservation, front office, payroll, accounting, restaurant management, human resources, and other systems are located (Cobanoglu & Cougias, 2003). At the user level, the user browse the hotel website to make reservations and to access the information about the hotel (Cobanoglu, & Demicco, 2007). This level of interaction in a network environment increases the vulnerability of computer systems to information and network security threats as any unencrypted information transmitted on Transmission Control Protocol/Internet Protocol (TCP/IP) can be sniffed by unauthorized people (Cobanoglu, & Demicco, 2007). The IT security problem exacerbates as most of the people in hotels who deal with computer systems (e.g., front office managers and accountants) are not IT experts and have only some familiarity with computer systems (Cobanoglu, 2005). Cyber risk is becoming an increasing issue for many organizations and businesses, including hospitality industry, as new advances in computing systems emerge, such as cloud computing, social media, and big data.

## Questions of this research and the purpose of study

The purpose of this study is to analyze cyber security and security practices of electronic information and network system, network threats, and techniques to prevent the cyber attacks in hotels. Helping the information technology directors and chief information officers (CIO) is the aim of this study to advance policy for security of electronic information in hotels and suggesting some techniques and tools to secure the computer networks.

The following research questions were created:

1. What methods, tools and techniques are currently used in hotels regarding computer network and system protection?

2. What are the current threats to computer network security in hotels?

3. What are the ways of handling security attacks in the hotel's computer networks?

4. What is the importance of network security in hotels?

5. Which methods hotels leverage to secure their websites for any data and financial transactions?

6. What criteria hotels have to consider in making a strong password for their computer networks and logins (computers and websites)?

## LITERATURE REVIEW

Information technology (IT) is the science and technology of using computers and other electronics to save and transmit information (Cobanoglu, & Demicco, 2007). This information can include variety of data, such as sales, customers, market, stocks and shares, accounts, airplane tickets, and booking (Cobanoglu, & Demicco, 2007).

Organizations that use IT need to tackle and administer electronic information safely and securely (Daler et al., 1989). The organization's administrative managers are responsible for the protection of the organization's assets and information (Cobanoglu, & Demicco, 2007).

Like other organizations, IT systems in hotels comprise of both software and hardware. The basic software in a hotel includes the property management system (PMS), point of sale system (POS), call accounting system (CAS), and accounting system. The basic hardware in a hotel include front desk computers, POS terminals, back office computers, printers, routers, switches, firewalls, and network cables (Cobanoglu, & Demicco, 2007). The front and back office computers, POS terminals, and printers are connected to routers and switches with network cables that enable communication between these devices (Cobanoglu, & Demicco, 2007). Routers also connect the hotel network with other networks and the Internet (Cobanoglu, & Demicco, 2007). The firewalls protect the hotel network from outside attacks (Cobanoglu, & Demicco, 2007). The hotel's local area network (LAN) typically consists of devices within the hotel's premises (Cobanoglu, & Demicco, 2007).

*Hospitality Industry and its Computer System*

Most of the computing systems are connecty computer networks to enable variety of applications, such as database sharing and shared storage (Cobanoglu, & Demicco, 2007). The hospitality industry uses the computer networks to manage reservations while avoiding duplex reservations for the same date and time (Cobanoglu, & Demicco, 2007). The hospitality industry uses two main types of computer networks: local area network (LAN) and wide area network (WAN) (Collins et al., 2003).

*Computer Network and Information Security*

Information security aims at maximizing the revenue of organizations and investments by minimizing the damage that could be caused by security attacks (Computer Fraud & Security, 2002). Most of the information security systems provide three main services: *confidentiality*, *integrity* and *availability* (ISO, 2000).

Organizations using IT are vulnerable to various security threats and attacks. The most common threats include viruses, inside attackers for network access, laptop theft, spoofing, unauthorized insider access, unauthorized outside attack, and denial of service attacks (Cobanoglu, & Demicco, 2007).

Computer crimes have always been there since the introduction of computers, however, the nature of attacks varies as the technology evolves (Beaulier Law Office, 2003). For instance, hacking or cracking is the crime of connecting to a computer system without permission (Cobanoglu, & Demicco, 2007). Most of the hacking attacks are aimed at obtaining confidential information (e.g., financial information of banking accounts, user accounts information) without authorization (Cobanoglu, & Demicco, 2007). Theft of technology occurs when an attacker consciously connects to a computer with intentions to steal technological information or secrets (Cobanoglu, & Demicco, 2007). Fraud happens when an attacker consciously connects to a computer with intentions of fraud or masquerades a legitimate user of the computer system (Cobanoglu, & Demicco, 2007). Hacking, theft of technology, and frauds are the most common security attacks whereas other security attacks are also possible (Cobanoglu, & Demicco, 2007).

*Information Security Tools and Techniques*

Information security aims at protecting valuable assets from disclosure or damage (Cobanoglu,

& Demicco, 2007). This protection can be attained through both technological and non-technological methods, such as physical security of assets, user identification and authentication, biometrics, and firewalls (Cobanoglu, & Demicco, 2007). We define some of the network and information security tools and techniques in the following:

1. *Digital IDs* are the electronic counterparts of driver's licenses, passports, and membership cards (Cobanoglu, & Demicco, 2007). Digital IDs often include a username and a password.

2. *Intrusion Detection System* is a system that analyzes the events happening in a computer system or a network for detecting intrusions or attacks (Judge, 2003). An intrusion can be defined as an effort to circumvent security services employed by the system, such as confidentiality and integrity (Cobanoglu, & Demicco, 2007). Many times intrusions are aimed at making the computer system unavailable (denial of service attack). Intrusions can be caused by various means: (1) attackers connecting to the systems from the Internet or outside networks; (2) authorized users of the systems who try to obtain additional privileges for which they are not authorized; and (3) authorized users who misuse and abuse the privileges given to them (Cobanoglu, & Demicco, 2007).

3. *Physical Security* refers to keeping the networking and computing equipment in a secure physical environment (Cobanoglu, & Demicco, 2007).

4. *Firewall* can be a hardware, a software or combination of hardware and software equipment to monitor the traffic between two or more computer networks (CERT, 2005). The firewall can also block particular malicious packets trying to enter a computer network.

5. *Encryption* is the process of hiding the information by making the information transformed in

a way that is impossible or very hard to understand (Cobanoglu, & Demicco, 2007).

6. *Biometrics* is technology of authenticating a user based on physical or behavioral characteristics, such as finger prints, voice recognition, and retina or iris identification. Biometric technology is one of the most effective methods of identity verification (Bilgihan, Karadag, Cobanoglu, Okumus, n.d.). The biometric systems measure the physical characteristics of an individual, and compare them with the recorded characteristics to verify the user's identity (Flink, 2002).

7. *Access Control* are techniques of restricting usage of system resources to authorized users/processes (Cobanoglu, & Demicco, 2007).

8. "*Vulnerability Assessment Scan* is a software that scans/examines the system for possible vulnerabilities" (i.e., weaknesses) and inform the system administrator of these vulnerabilities so that system can be safeguarded against these weaknesses (Cobanoglu, & Demicco, 2007).

Attackers can be from both inside and outside an organization (Cobanoglu, & Demicco, 2007). Especially in hospitality and tourism industry where turnover rate is very high, the possibility of inside attackers is higher in comparison to other industries (Cobanoglu, & Demicco, 2007). Consequently, some organizations, such as Burger King Corporation, take measurements to prevent inside attackers, and provide infrastructure to ensure only a single sign-on by an employee (Liddle, 2003). In this case, only one record needs to be expunged from the system in case of an employee's termination or resignation from an employee so that the former employee will not have any access to the system (Cobanoglu, & Demicco, 2007).

*Cyber Risks and Challenges in Hospitality Industry*

There are five major risks and challenges that hotels have faced during 2015 and they still continue, according to Hiller (January 23, 2015). These five challenges are: " 1. Identity theft leading to credit card fraud which has caused a lot of data breaches and information stealing from hotel's network systems, 2. Silent invasions and Cyber-crime attacks that are powerful tactics from next generation criminals, During 2015, there are lots of cyber criminals who have targeted and attacked the hotels' Wi-Fi and get the guests' personal information as well as their passwords. 3. Unfortunately there are no security audit cycles in majority of the hotels and this issue will put the situation of the investors and the guests in a highly risk. 4. Physical crimes like terrorism that put the hotels in challenge and it can be more seen in South Asia and Middle East. 5. Loss of competitive advantage and image as well as lots of negative words of mouth is other challenges that hotels have faced due to the cybersecurity attacks" (Hiller, 2015).

One of the unique features of hospitality industry is being a place for their customers' comfort and confidence; thus it is basically rely on this reputation and abusing the customers' information doesn't make any sense in this regard.

Unfortunately the reality shows that this confidence and reputation can't be achieved easily due to the estimation of annual cost of cyber crime that can affect the global economy as much as $375 billion to $575 billion and these numbers are still growing according to Butler ((January 15, 2016), due to greater technology available in the market for cyber attackers that makes the hotels' system more and more vulnerable.

In this regard, hotels are also not completely innocent and without fault; because they can do a lot to prevent these cyber attacks; for example in the case of data breach which happened to Wyndham hotels, after the investigation, many mistakes were revealed which was made by the

IT managers of whoever was in charge of taking care of information technology system and computers. Some of those mistakes according to Butler ((January 15, 2016), are mentioned below:

- "failed to use readily available security measures, such as firewalls
- stored credit card information in clear text
- failed to implement reasonable information security procedures prior to connecting local computer networks to corporate-level networks
- failed to address known security vulnerabilities on servers
- used default user names and passwords for access to servers
- failed to require employees to use complex user IDs and passwords to access company servers
- failed to inventory computers to appropriately manage the network
- failed to maintain reasonable security measures to monitor unauthorized computer access
- failed to conduct security investigations
- failed to reasonably limit third-party access to company networks and computers"

With what was mentioned above, we will understand that some of these cyber attacks can be prevented by hoteliers; however these cyber crimes will not come to the end here since cyber attackers will find new ways to penetrate customers' information and take advantage of them. One of the most famous ways to get into customers' account information is opening a new website and pretend to be the hotel that the customer looks for by using the same brand name, logo, map, picture and all other information that belong to the real hotel (Mest, October 6,2015). According to Mest (October 6,2015) this process is called fake booking which The American Hotel & Lodging Association reported that, each minute there are about 480 fake online hotel booking website are creating and it is an amazing place for cyber criminals to take advantage.

Data breaches in the world of business continuously increase and they remind us of how having a secure system, taking action on cyber risks and implementing different security tools are critical and necessary for any business (MASTER, October 8, 2015). According to MASTER (October 8, 2015), not having enough knowledge and in most cases (%80) not having any knowledge at all is the biggest mistake, which is made by business owners. Therefor, as long as businesses doesn't have any idea about how to secure their business, hackers and cyber attackers will have more opportunity to take advantage of this issue which is like an easy treasure for them to gain. In addition, hackers and attackers are not only able to abuse the computer system of the hotels by using different type of phishing email, viruses and etc., but also they are able to attack and take advantage of Wi-Fi in the hotels (Clark, April 13, 2015). Most of the hotels now a days offer free Wi-Fi to their guests and the guests will access to the same network all over the hotel such as lobby, convention center, dinning room and all other places within the hotel (Clark, April 13, 2015). This is one of the advantages of the hotels however it can seriously cost a lot for the hotels (Clark, April 13, 2015). According to Clark (April 13, 2015), in some of the cases which are reported, the router that was infected by hackers plus the ANTLabs InnGate ("Specially designed for the hospitality sector, the ANTlabs InnGate meets the High Speed Internet Access (HSIA) needs of hotels and service apartments, while allowing hoteliers to roll out free HSIA WiFi that pays for itself."(InnGate3, n.d.)), were associated with the hotel's property management system so that hackers and cyber attackers could easily access to whatever they want to know about the hotels such as customers information, database, property payroll system and a lot more (InnGate3, n.d.)). Furthermore, by taking advantage of hotels' Wi-Fi, hackers offer "updates" for software that are famous such as Adobe Reader or Flash Player so that the users won't hesitate to update their software and then those updates contain malware that criminals use to get all the usernames, passwords or any other important information from users' computer or smartphone (Noone, August 5, 2015). Based on this fact, Noone (August 5, 2015) suggested a couple of tips and advices for hotel customers while being in travel and especially in the hotel such as not using online banking on public computers and public Wi-Fi, not access or in the case of emergency, access your email inbox by creating a throwaway email address, prevent your computer or smartphone from automatically connection to any unknown Wi-Fi, instead of saving sensitive information on the laptop or smartphone, use remote desktops.

Interestingly, in most of the cases, the software that is used by hackers and cyber criminals, is not a new software and it can be even up to 10-year old software and unfortunately due to ignorance of the hotels and not updating their system, even sometimes this software can be taken advantage of by previous contractors that hotels hired for their information technology support (Greenberg, February 2, 2010).

Another issue, which is related to hotels cyber security and customers' information safety, is the matter of loyalty programs. According to Friedland (September 10, 2015), in several of studies that have done so far, most of the loyal customers were not ready to consider to be a loyal customer of the same hotel again after the hotel experienced breach. For any brand, loyal customers are gifts and very valuable assets that the company should always try to keep (Friedland, September 10, 2015). In addition, it is very cheaper to keep a loyal customer than to make new customers for the brand. Thus, all hotels must do their best to keep them satisfied by securing their information and protect them from cyber attackers (Friedland, September 10, 2015). The possibility of taking advantage from loyal customers' information is much more than any other customers since due to lots of loyalty programs, loyal customers' profile have more details and information in itself which makes it more interesting for cyber attackers to be curious about and try to take advantage of (Friedland, September 10, 2015); In this case the responsibility of hoteliers is more about loyal customers while protecting their information and secure their experience (Friedland, September 10, 2015). Regarding that, according to Friedland (September 10, 2015), IT manager or GM of the hotels can do a couple of things to strengthen then security regarding protecting loyal customers' information and data:

1. "Giving them information about the possibility of being hacked by cyber attackers and alarm them to regularly change their password and avoid having the same password for several websites. Also inform them to check their account activity more often and reward them for being securely active as well as practicing secure ways to use their account.
2. Sending them an automatic email and notify them in case of changing the password or login to their account, so in this case if they haven't logged in to their account and received an email regarding that, they can immediately be aware of abuse and report it.
3. Empowering the system to have a two-factor authentication so that for logging into accounts, people require to submit the security code that they will receive on the same email address or phone number that they provided before while they were singing up to the website. Therefore, in case of abusing, cyber attackers will not be able to access to the account as long as they don't know the email address or phone number that the account was registered. Even if they know the phone number or the email address, they will not have the actual phone or the password of the email address to enter and get the pass code."

*DSS PCI Compliance level*

In addition to all that, any organization and industry that works with credit card must accept the risks and cost of fraud in doing business (How credit-card transactions just got dicier for hotels, October 29, 2015). Payment Card Industry Data Security Standards (PCI DSS) helps a lot to secure the credit card, debit card and cash transactions and protect them from fraud and misuse by determining some policies and procedure (How credit-card transactions just got dicier for hotels, October 29, 2015). The PCI DSS was created on 2004 jointly by the four major credit card companies, which are Visa, MasterCard, Discover and American Express (Rouse, n.d.). According to Amazon Web Services (n.d.), all organizations that process the credit card information are require to be certified regardless of their deployment model. Having a PCL DSS certification requires the organizations to develop and maintain a secure network. It also requires protecting the cardholder data and information as well as maintaining a vulnerability management program. Furthermore, organizations require to implement the strong security measures and frequently test, control and monitor networks and finally maintain an information security policy (Amazon Web Services, n.d.)

There are different levels of PCI compliance that all merchants will fall into, according to Visa transaction volume over a 12-month period. This transaction is determined based on the aggregate number of Visa transactions (inclusive of credit, debit and prepaid) from a merchant Doing Business As (ComplianceGuide, n.d.). There are four different merchant levels that any merchants based on the categories that are described below and regardless of their acceptance channel, should have (ComplianceGuide, n.d.).

"**Level 1:** Any merchant that process more than 6M Visa transactions per year.

**Level 2:** Any merchant that process 1M to 6M Visa transactions per year.

**Level 3:** Any merchant that process 20,000 to 1M Visa e-commerce transactions per year.

**Level 4:** Any merchant that process fewer than 20,000 Visa e-commerce transactions per year, and all other merchants that process up to 1M Visa transactions per year."

*Cyber Risks and Cyber Attack Prevention Methods*

We should know that none of the security software, antiviruses, and other tools can % 100 guarantees to prevent hotels and any other business from cyber attacks (Accenture operations, n.d.). On the other hand, securing and limiting the system too much may cause other problems such as preventing the customers to access the information they might need; thus, hotels and other organizations should manage the risk in cyber defense which according to (Accenture operations, n.d.) requires a meaningful and understandable operational model which causes balance in security implementation and operation as well as using the newest technology available plus testing of security posture and very strong feedback structure (Accenture operations, n.d.). In Accenture operations website (n.d.), for cyber defense excellence, there are three steps available, which are: 1. Prepare and protect 2. Defend and detect and 3. Respond and recover.

Now we will go to the details of each step to see what each step suppose to do to secure the business as much as it is able to (Accenture operations, n.d.).

1. PREPARE AND PROTECT: this step will give us a big view of the security performance in support of businesses. It makes us aware of the threat intelligence existence and will make us ready to manage our business vulnerability. In this between, experts will need to have forward thinking capabilities to help scale activities and on the other hand IT strategies will be designed based on great understanding of assets, data sets, technical and business functions.

2. DEFEND AND DETECT: in this step which is very critical, the forward thinking capabilities should effectively help the scale activities so that operational monitoring and controlling capabilities be able to analyze the security in an advance level which focuses on visualization to understand and identify the anomalies and suspicious activities.

3. RESPOND AND RECOVER: in the last step which focuses on intelligent incident response, some active defense strategies will be take into action which requires the security incidence management. This step is the art of the platforms to catch the hackers and attackers or any other threat to the business.

In general, Cyber attacks can occur in any four forms below (Bamrara, 2015):
1. "The intruder may unofficially access to the network
2. The intruder may destroy, otherwise or corrupt the alter data
3. The intruder may fake the permission from user system and enter to the system
4. The intruder may implement some malicious procedure to fail, hang or reboot the system while can not access the system"

The modern system of security really makes it difficult for hackers and cyber criminals to attack and access the network system however if the system is not configured in a good and proper way or the update patches are not installed then cyber criminals may crack to the system while using security loophole (Bamrara, 2015).

According to Lavelle (Jan 4, 2016), most of the cyber attacks occur by sending phishing emails, hacking weak passwords or some other attacks, which target the vulnerabilities of the web applications. Thus, a lot of endpoint security vendors offer the anti-phishing solutions (Lavelle, 2016).

One of the tools to prevent data breach attacks is Web Application Firewall (WAF) that hotels can take advantage of. WAF solutions can be accessed through market and they are able to utilize as a cloud-based service and an on-premise appliance (Lavelle, 2016). WAF solutions are also useful to detect and prevent the data to be revealed, therefore the attackers that target the credit card database will not able to get those information since the WAF solutions will detect and block the database (Lavelle, 2016).

One of the best ways that hotels can use in order to secure themselves as well as customers' information and protect their data and also cover their loss in the case of data breach is having cyber security insurance. According to Butler (January 15, 2016), cyber security insurance must be a concentration for any hotel owners. Hotel owners have to know that data breaches and cyber claims are not included in general liabilities policy; therefore this fact makes it even more important for hotel owners to think about and getting insurance for any cyber claims in case of cyber attack. Cyber security insurers will cover both first and third party in the case of cyber attack and data loses. The third party can be both customers and government or any regulatory agencies.

*Learning about Digital Certificates*

Digital certificates are used to bind a message to the owner/generator of the message (Chapter 6 - Digital Certificates, n.d.). This binding is established by the use of private keys, that is, the owner of the message signs a message with his private key (only the owner has its private key) (Chapter 6 - Digital Certificates, n.d.). The verification is done with the public key of the message owner/generator. Digital certifications are important in cybersecurity as they provide non-repudiation service (Chapter 6 - Digital Certificates, n.d.). In hospitality industry, digital certifcates are also crucial to prevent fraud from customers or hotel/restaurant owners as false claims from either can be legally challenged and the truth be established by using digital certificates (Chapter 6 - Digital Certificates, n.d.). Digital certificates are integrated in all modern web browsers so that the identity of the people and organizations can be verified, and the integrity of the content be preserved (Chapter 6 - Digital Certificates, n.d.). The following information (at least) is contained in each digital certificate (Chapter 6 - Digital Certificates, n.d.):

- "Owner's public key
- Owner's name or alias
- Certificate's expiration date
- Certificate's serial number
- Name of the organization that issued the certificate
- Digital signature of the organization that issued the certificate "

Certificates are authenticated, issued, and managed by a trusted third party called a *certification authority* (CA). The CA must provide following services: (Chapter 6 - Digital Certificates, n.d.)

- "Security related technology, such as security protocols and standards
- Infrastructure, including secure facilities, backup systems, and customer support
- A legally binding framework for managing subscriber activities and resolving disputes"

Digital certificates can be used for the following: (Chapter 6 - Digital Certificates, n.d.)

- "To verify the identity of clients and servers on the web
- To enable secure communication between clients and servers by encryption (e.g., secure Internet email communication)
- To associate a digital signature with an executable code that users can download from the web to verify the source and integrity of the executable code "

Clients and servers exchange digital certificates using a secure transmission protocol, such as secure sockets layer (SSL) or transport layer security (TLS). (Chapter 6 - Digital Certificates, n.d.)

*How to know if a Website is Secure*

Before providing any confidential information to a website, one should ensure that the website is secure. The following tips help in identifying the security of a website: (Woodfield, November 26,2014).

**1. Check the SSL Certificate**

The website address, also known as the website's Uniform Resource Locator (URL), provides information about the security of a website. If the URL begins with "https" instead of "http", it implies that the website uses SSL and is secure (Woodfield, November 26,2014). The website hosting company has to go through a validation process to obtain an SSL Certificate (Woodfield, November 26,2014). There are a few different validation levels and some of them are more difficult to obtain than the others (Woodfield, November 26,2014). The two most commonly used validation levels are: (a) Domain Validation (DV); and (b) Extended Validation (EV). The DV is the lowest level of validation that simply verifies the ownership of the domain and not the legitimacy of the organization requesting the certificate (Woodfield, November 26,2014). For example, if someone purchases a domain "amaz0n.com" and requests a certificate for this domain, he/she would get the certificate because of the domain's ownership (Woodfield, November 26,2014). The EV is the highest validation level (also the most safest and expensive validation level) that requires not only the company requesting the certificate to prove its identity but also verifies the company's legitimacy as a business (Woodfield, November 26,2014). A website having an EV certificate can be recognized by looking at the address bar (Woodfield, November 26,2014). Browsers depict a green address bar with a padlock icon for websites with EV certificates, as shown in the picture below (Woodfield, November 26,2014).

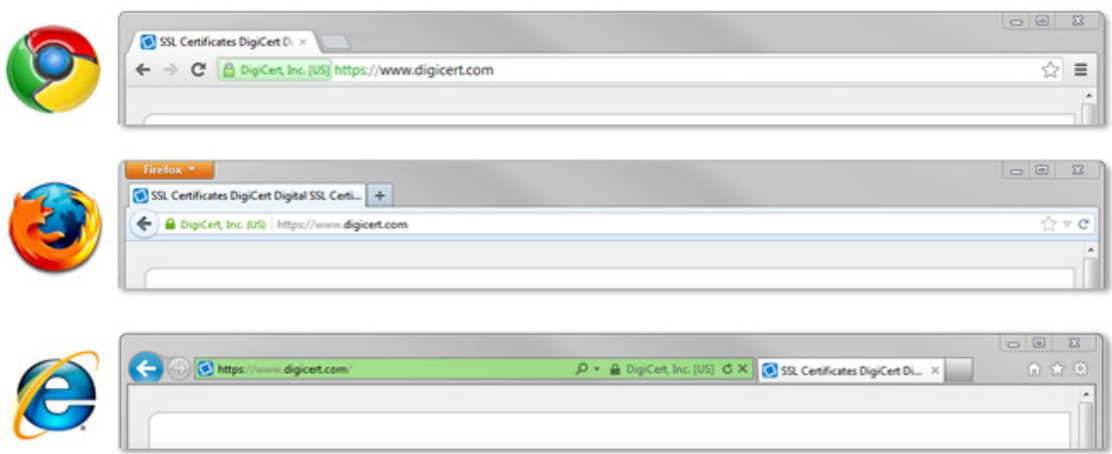

Fig. Websites with Extended Validation (EV) Certificate

**2. Look at the Website Domain**

Cyber attackers sometimes make websites that mimic existing websites in order to trick people to enter their confidential information (e.g., credit card numbers, login information) on the phishing website. For example, assume a cyber attacker purchases the domain "amaz0n.com" and sets up a website at "amaz0n.com" that resembles the amazon.com website (Woodfield, November 26,2014). The attacker purchases a DV certificate for the website and attempts to trick users (by using phishing emails or other methods) to buy items or log into their accounts on the mimic phishing site (Woodfield, November 26,2014). These phishing attacks can be avoided by always looking at the website domain to ensure that one is connected to the authenticated website and not a phishing website (Woodfield, November 26,2014).

### 3. Look for Signs that the Company is Real

One should always look for signs to identify if the company represented on the website is real or not (Woodfield, November 26,2014). The following signs can help in the identification of a legitimate company: (Woodfield, November 26,2014).

*Physical address and phone number* – Listing of a physical address and a phone number on the website increases the chances that the business is legitimate because reputable companies often list their information so that a customer can contact them if there is a problem (Woodfield, November 26,2014).

*Return policy* – Reputable e-commerce sites should list their return and shipping policies. If one cannot find the return and/or shipping policy of a company, one should refrain from purchasing on such websites (Woodfield, November 26,2014).

*Prices are too low to believe* – If a website offers a product at a price that is incredibly lower than the actual product price, this could be an indication that the website is selling a counterfeit product or one may not get anything at all after purchasing from such a website (Woodfield, November 26,2014).

*Privacy statement* – Reputable sites should list their privacy statement, that is, how the website protects customer information and whether the website shares the customer information with third parties (Woodfield, November 26,2014). One should make sure a site has a privacy statement and should read the privacy statement before making a purchase (Woodfield, November 26,2014).

**Examples of Cyber Attacks in Hotels**

In 2014 and 2015, we will see a large number of hotels that were affected by cyber attackers, which has never seen before in the history of hotel industry. Below are some stories about these hotels that were exposed and had caused a lot of guests' personal information and their bank accounts to be revealed to attackers:

1. According to Hertzfeld (September 3rd, 2015), There was a data breach that was reported at a Brunswick Maine and Hotel, which possibly could affect 2,600 guest information who had stayed in the hotel between November $29^{th}$ to July $21^{st}$ Based on company estimation. The Brunswick Hotel and Tavern management said, malware was placed on their front desk computer and the CFO (Dan Flaherty) believed that through a scam phishing email, which pretended to be a message from a guest, the malware was installed. Malware was designed in a way to avoid demodulation from antivirus software. According to a letter from one of their customer, which was posted online by Vermont Office of the Attorney General, the malware's aim was to capture and permit remote access to payment card information. Finally the hotel got some help from the cyber security and investigation company to search more about the case as well as make the whole system of the hotel more secure and increase the online safety for their customers.

2. On March 2015, the Mandarin Oriental Hotels experienced the data breach (Lavelle, 2016). The breach was designed to get the credit card and debit card information of those customers who used spa, beverages, guest rooms, dining room and other product and services. It is also infected the sales system of a couple of properties (Petri, Dec 08, 2015).

3. From November 18 to December 5, 2014 and from April 21 to July 27, 2015 Hilton Worldwide hotels faced the data breach, credit and data theft as well as some of its point-of-sale computer system. Fortunately, the hackers were not able to get the name of the customers as well as their addresses and personal identification numbers, however they had accessed to the credit card numbers plus security codes and expiration dates. Hilton got help from a third-party experts, law enforcement and payment card companies to investigate the breach and also told its customers to see if there is any irregular activity in their accounts (AFP, Nov 25, 2015).

4. On November 2015, Four days before attack to Hilton Hotels, Starwood Hotel Group got attack and that caused data breach. Hackers targeted the payment system and potentially leaking customer credit card data. Starwood also got help from forensic experts so that could recover the system in a couple of the restaurants as well as gift shops and point of sale system of the hotel (AFP, Nov 25, 2015). The company was also gave the list of the affected hotels which included the Sheraton New York Times Square, the Westin New York Grand Central New York and the St.Regis Bal Harbour Resort in Florida (Starwood says malware, November 24, 2015).

5. Hyatt Hotels also faced data breach on December 2015 (Lavelle, 2016). It did not talk about the type of the malware that was used to attack the data however the hotel asked the customers to check their account activity and also told them that they can feel free to using payment cards at Hyatt hotels since they are making their security system very strong to protect their customers (Rukmangadhan & Finkle, December 24, 2015). This data breach is a big warning to Hyatt hotels because it makes all their 627 properties in 52 countries vulnerable to the cyber criminals (Champagne, January 6, 2016).

6. And finally, on October 2015, Trump Hotel Collection confirmed the data breach and it alarmed the customers who used their credit card in the hotels between May 19, 2014 and June 2, 2015. This data breach happened in several Trump locations such as SoHo New York, Trump International Hotel and Tower Las Vegas, Trump International Waikiki and Trump International Chicago (Petri, Dec 08, 2015).

## *METHODOLOGY*

This study is completely qualitative and the case study is targeting the hotels in Reno, Nevada. Based on that, five hotels are chosen without any specific consideration or hotels' features. The researcher decided to have an interview with a targeted sample for the following reasons: 1) interview is one of the most effective tools and techniques to collect data; 2) the simplicity, practicality, flexibility and interactive nature of the interview provides an opportunity to hear the interviewee's perspective regarding the subject and research.

The interviews for this study were conducted either face-to-face or via phone for data collection. The interviewees included hotels' front desk employees, hotels' guests, and hotels' assistant of general managers (GMs) or IT managers depending whether the hotel employed an IT manager or not. The participants were selected based on their knowledge about the issue or their ability to give productive, useful and effective information about the topic that can be contributed to the public knowledge, such as public knowledge about cyber security.

In each hotel, ten guests were interviewed without giving any consideration of whether the interviewee was a loyal customer or not. Also two Front Desk employees were interviewed in each hotel. In three hotels out of the five that had an IT manager, the interviews were done with them; however, the other three hotels, which didn't have IT managers, the interviews were done with assistant of GMs.

Since this study is not a human subject study, the researcher didn't get IRB for the study; however, all interviewees were informed about the confidentiality of the information they had given. Each interview with guests took about 7-10 minutes and with employees it took about 5 minutes. Interviews with IT managers and Assistant of general managers took about 15 minutes each; however unfortunately the researcher couldn't get enough information that the study seeks due to the confidentiality of the hotels' security information. IT managers and Assistant of GMs informed the researcher that they were not allowed to reveal anything about their network security system and their implications regarding that because it was against their policy and the policy of the third party company that they work with to reveal hotel's security-related information. Therefore, some of the interview questions couldn't be answered.

The interview questions were made based on some of the previous articles and works that have done so far regarding network security issues in hotels and hospitality industry. The interview questions did not ask of personal information about the interviewees and the participation in this interview was absolutely volunteer. Also, the interviewees were free not to answer any of the questions they did not want to. The answers to the interview questions were recorded on the piece of paper and later on were typed and added to the research project.

The interviews questions from all three groups (guests, front desk employees and assistant of GMs or IT managers) in each hotel can be seen in Table 1, and the answers are available in the "Findings and Results" section.

The data analysis of this study is based on the qualitative nature of the study. Since it is a case study research, the data analysis will be done with pattern matching and explanation building in conclusion section. These two methods of data analysis are introduced and defined by Yin (1994).

Table 1: Interview Questions

| Question's Number | Interview questions for Assistant of GM/ IT managers | Interview questions for Front Desk employees | Interview questions for hotel's customer |
|---|---|---|---|
|  |  |  |  |

| # | | | |
|---|---|---|---|
| 1 | What are the main property (hotel) features, such as star rating, number of rooms, services they offer (e.g., spa, swimming, Wi-Fi, parking, restaurants, etc.)? | Does the hotel have any training program for employees to inform them the importance of computer security and to use the network securely? | Have you ever experienced any fake website for booking a hotel? |
| 2 | Does the hotel employee a dedicated IT manager or this position is handled by the hotel's GM? | If the previous question's answer is yes, Then to what extend and how they trained you? | Have you ever been informed the hotel you used to stay in, experienced a data breach that caused you to see some unusual activities on your credit/debit card account? |
| 3 | What are the current practices used by hotels to protect their computer networks? | How often is there is any, do you receive phishing emails? | As a customer of the hotel, if you heard the hotel computer system is attacked, will you still choose this hotel to stay in? |
| 4 | What kind of computer security threats (if any) the hotel has experienced in past five years? | How do you recognize if an email is phishing or not? | If you are a loyal customer of this hotel, how often do you change your password of your loyalty account and check its activities? |
| 5 | Has the hotel reported publicly any computer security threat (if any) experienced in the past? | Have you ever noticed if the hotel has experienced any data breach? | What are your criteria to choose a password for your accounts? |
| 6 | Which methods the hotel leverages to secure their websites for any data and financial transactions? | Have you ever received any complain from guests about fake website reservation? | Do you pay attention to the website security of the hotel you are going to book and pay online? If yes, then what do you pay attention to? |
| 7 | What criteria the hotel considers in making a strong password for their computer networks and logins (computers and websites)? | | |

| 8 | Does the hotel have some contract/arrangement with an IT firm to maintain as well as to handle the security of their cyber infrastructure? | | |
| --- | --- | --- | --- |
| 9 | What measures hotel takes to secure the confidential data from potential internal threats (e.g., rogue employees)? | | |
| 10 | Does the hotel have any plan to incorporate more security measures in their cyber infrastructure? | | |
| 11 | What is the level of your compliance for PCI DSS? Do you believe 100% compliance is possible? Why? Why not? | | |
| 12 | What are your biggest security challenges? | | |

## *FINDINGS AND RESULTS*

The results and findings of the research is shown below:

**Hotel 1**

This hotel hasn't had an IT manager, so the interview was done with the assistant of GM. This hotel is a 2-star hotel with 59 smoke-free guestrooms, Free airport shuttle, Fitness center, Business center, 24-hour front desk Coffee/tea in the lobby, Air conditioning, Computer station, Self-serve laundry, Meeting rooms, Multilingual staff and Free newspapers in the lobby. At the moment the hotel uses Anti-virus software to protect its system; however the hotel's computer security is very low and it doesn't require any password for guests to log into the hotel's Wi-Fi, which makes the computer system very vulnerable. The Assistant of GM claimed that they have not experienced any attack in last five years. Based on that the hotel has not reported anything to public about any cyber attack. The website of the hotel is certified and secure for any financial transactions. The hotel currently has decided to make a contract with an IT company, however they do not have any specific IT company at the moment to work with. For now in case of facing any problem, they acquire services of IT professionals from various companies. The hotel tries to have a strong password to access the computer system by using small and capital letters as well as symbols and numbers in creating passwords. Unfortunately the hotel doesn't have any specific method to protect the confidential data from potential internal attackers, however the front desk employees are not able to see some information that managers can see. The assistant of GM believes that there is no %100 in security.

The result after interview with front desk employees shows that the hotel does not have any specific training program regarding cyber security, however both employees claimed that they have an idea about phishing emails, so they do not open all emails that they receive. They mentioned they get phishing emails very often and lots of other emails that go to spam section of the email and based on the content of email and the format they can recognize if this is a phishing email or not. They also mentioned they never heard about data breach in this hotel and they barely received complains about the fake websites and money transactions.

Among 10 guests in this hotel, only one of them said that he has experienced a fake website while he was booking a hotel for his business purposes 2 years ago and he didn't recognize the issue until the time he went for check-in and front desk employee told him that his name is not there. He also mentioned that the front desk employee already knew about the fake website since they had another customer who had claimed the same thing. All of the guests responded no to the second question regarding being informed about any data breach in the hotel they have stayed; however one of them mentioned that after coming back from his travel, he observed some unusual credit card activities, so he called the bank and informed them and get his refund. All guests responded no regarding staying in a hotel that has experienced any data breach. Among ten guests, three of them were loyal customers however they said they never change their passwords and their criteria for choosing password is adding numbers, symbols and small and capital letters to the password and make it as strong as possible. Among all guests, only three of them said they pay attention to the website security while paying online.

**Hotel 2**
This hotel has an IT manager and it comes under 4-star hotels. The amenities of the hotel include: 1990 guestrooms, Casino, 10 restaurants, Full-service spa, Nightclub, Breakfast, Free airport shuttle, Seasonal outdoor pool, Ski storage, Health club, 24-hour business center and Limo/town car service. The interview was done with the IT manager. Due to the security of the hotel, the IT manager was not able to provide ant information regarding their current security practices, their experiences in last 5 years, their PCI DSS compliance level, their challenges and the methods they use to prevent internal attacks. The IT manager mentioned that the hotel has not reported any cyber attack to public in last five years. The website of the hotel is completely secure for any personal and credit card information as well as financial transaction. The hotel also has annual contract with an IT company to maintain and handle the security of the cyber infrastructure. The hotel has a plan to get cyber insurance as well in the near future. The IT manager believed that there is no %100 security in anything and IT managers are always one step behind the hackers and cyber attackers.

The result after interview with front desk employees shows that the hotel does not have any specific training program regarding cyber security, however the IT manager frequently check the system to see if everything goes well. They mentioned they get phishing emails almost everyday and most of the emails in spam section are phishing. Based on the content of email and the format they can recognize if this is a phishing email or not. They also mentioned they never heard about data breach in this hotel and they received complains a couple of times about the fake websites and money transactions.

Among 10 guests in this hotel, none of them said that he has experienced a fake website for booking a hotel. All of the guests responded no to the second question regarding being informed about any data breach in the hotel they have stayed. All guests responded no regarding staying in a hotel that has experienced any data breach. Among ten guests, six of them were loyal customers however only two of them mentioned that they change their passwords every six months. These two guests were a couple who had studied and worked in IT area. Seven out of ten guests mentioned that their criteria for choosing password is adding numbers, symbols and small and capital letters to the password and make it as strong as possible and the rest of the guests mentioned that they use the same password for all their online accounts. Among all guests, seven of them said they pay attention to the website security while paying online.

**Hotel 3**

This is a 4-star hotel that has an IT manager plus1623 guestrooms, Casino, 10 restaurants and 14 bars/lounges, Full-service spa, 2 outdoor pools, Nightclub, Breakfast, Internet in the lobby, Free airport shuttle, Ski storage, Fitness center and 24-hour business center. The interview was done with the IT manager. Due to the security of the hotel, the IT manager was not able to provide ant information regarding their current security practices, their experiences in last 5 years, their PCI DSS compliance level, their challenges and the methods they use to prevent internal attacks. The IT manager mentioned that the hotel has not reported any cyber attack to public in last five years.

The website of the hotel is completely secure for any personal and credit card information as well as financial transaction. The hotel also has annual contract with an IT company to maintain and handle the security of the cyber infrastructure. The hotel does not have cyber insurance yet. The IT manager believed that there is no %100 security.

The result after interview with front desk employees shows that the hotel does not have any specific training program regarding cyber security, however the IT manager frequently check the system to see if everything goes well. They mentioned they are not responsible to check the emails and response and they have another department for that. In addition, they also mentioned they never heard about data breach in this hotel and they received complains a couple of times about the fake websites and money transactions.

Among 10 guests in this hotel, two of them said that they have experienced a fake website for booking a hotel an one of them said that after paying and booking the hotel room, her son had recognized the website was fake due to the logo image. All of the guests responded no to the second question regarding being informed about any data breach in the hotel they have stayed. Nine guests responded no regarding staying in a hotel that has experienced any data breach and only one said if they informed the guests about it and they guarantee the security now, she will still stay at that hotel. Among ten guests, five of them were loyal customers however none of them change their password regularly and two of them said in the case of observing suspicious activities, they would change the password. Five out of ten guests mentioned that their criteria for choosing password is adding numbers, symbols and small and capital letters to the password and make it as strong as possible and the rest of the guests mentioned that they use the same password for all their online accounts. Among all guests, six of them said they pay attention to the website security while paying online.

**Hotel 4**

This hotel is a 4-star hotel that has an IT manager and the interview was done with him. The amenities of the hotel include: 824 guestrooms, Casino, 8 restaurants, Full-service spa, Indoor pool, Free airport shuttle, Ski storage, Health club, Conference center, Limo/town car service, 24-hour front desk and Air conditioning. Due to the security of the hotel, the IT manager was not able to provide ant information regarding their current security practices, their experiences in last 5 years, their PCI DSS compliance level, their challenges and the methods they use to prevent internal attacks. The IT manager mentioned that the hotel has not reported any cyber attack to public in last five years. The website of the hotel is completely secure for any personal and credit card information as well as financial transaction. The hotel also has annual contract with an IT company to maintain and handle the security of the cyber infrastructure. The hotel does not have cyber insurance yet. The IT manager believed that there is no %100 security.

The result after interview with front desk employees shows that the hotel has a training program but not specifically for regarding cyber security, however the IT manager frequently check the system to see if everything goes well and also inform them not to answer security questions and not to provide private information of the hotel to anyone either on phone or email. They mentioned they get phishing emails a lot and most of the emails in spam section are phishing. They mentioned sometimes they are not able to recognized if the email is phishing or not because they look very normal. They also mentioned they never heard about data breach in this hotel and they received complains a three times in last year about the fake websites and money transactions.

Among 10 guests in this hotel, none of them said that they have experienced a fake website for booking a hotel. All of the guests responded no to the second question regarding being informed about any data breach in the hotel they have stayed. All of the guests responded no regarding staying in a hotel that has experienced any data breach. Among ten guests, two of them were loyal customers; however only three of them change their password regularly and six of them said in the case of observing suspicious activities, they would change the password. Eight out of ten guests mentioned that their criteria for choosing password is adding numbers, symbols and small and capital letters to the password and make it as strong as possible and the rest of the guests mentioned that they use the same password for all their online accounts. Among all guests, eight of them said they pay attention to the website security while paying online.

**Hotel 5**

This 3-star hotel doesn't have an IT manager and the interview was done with the assistant of GM. The amenities of the hotel include: 260 guestrooms, Casino, 2 restaurants, Breakfast, Free airport shuttle, Seasonal outdoor pool, 24-hour business center, 24-hour front desk, Air conditioning, Front desk safe, ATM/banking services and Laundry service. At the moment the hotel uses Anti-virus software, physical security and software firewall to protect its computer system. The password for log into the Wi-Fi for guests is changes daily; however on the same day the password is the same for all guests. The Assistant of GM was not willing to provide information about the hotel cyber attack, however she mentioned that they experienced once and it was not a major one so the hotel has not reported anything to public regarding that. The website of the hotel is certified and secure for any financial transactions. The hotel currently has contract with an IT company and they come to check the system and update the security software on a regular bases. The hotel tries to have a strong password to access the computer system by using small and capital letters as well as symbols and numbers in creating passwords. Unfortunately the assistant of GM was not able to provide any information regarding the methods to protect the confidential data from potential internal attackers. The assistant of GM does not believe in %100 securities and instead believes in luck and fortune.

The result after interview with front desk employees shows that the hotel does not have any specific training program regarding cyber security. They mentioned they get phishing emails sometimes and mostly spam emails from unknown people. They also mentioned they never heard about data breach in this hotel and they have never received complains about the fake websites and money transactions.

Among 10 guests in this hotel, none of them said that they have experienced a fake website for booking a hotel. All of the guests responded no to the second question regarding being informed about any data breach in the hotel they have stayed. All of the guests responded no regarding staying in a hotel that has experienced any data breach. Among ten guests, one of them was loyal customer; however only two of them change their password regularly and eight of them said in the case of observing suspicious activities, they would change the password. Seven out of ten guests mentioned that their criteria for choosing password is adding numbers, symbols and small and capital letters to the password and make it as strong as possible and the rest of the guests mentioned that they use the same password for all their online accounts. Among all guests, eight of them said they pay attention to the website security while paying online.

# *CONCLUSION*

This study tries to explain the importance of the cyber security in hospitality industry. It also discusses about the tools and techniques to prevent cyber attacks. The findings and results of this study after interview with the front desk employees, guests and IT manager/ Assistant of GM in five different hotels that were chosen randomly, shows that not all the hotels in Reno, Nevada has IT manager or someone who dedicates to computer system and networks. Even some of the hotels do not have contracted with any specific IT company to refer when the face any challenge or problem and they call random IT professionals from different companies to fix their computers' problem. Also the current cyber security practices in some hotels are not enough at all to prevent the cyber attacks at all. In addition, due to the security and privacy of the hotel, the researcher was not able to understand whether the hotels have experienced any data breach in last five years or not and the cyber challenges that they are facing as well as their PCI compliance level and their criteria for creating the passwords; however the results show that all five hotels have not reported any data breach to public in last five years. Interestingly one of the hotels has a plan to sign a contact with an IT company very soon. Moreover, all five hotels' IT managers/ Assistant of GM do not believe in %100 security and they think this is not possible. On the other hand, the results of interview with front desk employees show that they barely receive any complain from guests about fake websites and there is no specific training program from the hotel for them regarding cyber security of the computer networks, while they recognize the phishing and suspicious emails based on their own knowledge about the content and outlook of the emails. They all also mentioned that they have never heard anything regarding data breach about the same hotel they work for.

And finally the results of the interview with hotels' guests show that nobody has ever heard about data breach in the hotel they stays while nobody wants to stay in a hotel that has experienced any type of data breach. Also more than half of the guests who were interviewed (32 out of 50), mentioned that they pay attention to the security of the website while booking the hotel online. In addition, more than half of the guests 27 out of 50) add symbols, numbers, capital and small letters to their passwords to make sure the password is strong enough and less than half of them change their password regularly, while most of them either use the same password for all their online account or change their passwords if they observe any suspicious and unusual activity.

So based on the findings and results and data analysis, we understand that hotels are very vulnerable and they definitely require a lot more security for their computer networks and system. While the hotels are very careless in paying attention to their network security, a lot of hackers and cyber attackers out there are waiting for the chance to get into the hotels' network system and access to the personal information of the guests without legal and official permission and authorization. One of the biggest reasons that cause cyber attacks is not having an IT manager or a company to monitor the computer network and system. Also the lack of knowledge in hotel employees and lack of training in cyber security makes the hotels more vulnerable. In addition, lack of public knowledge about cyber security causes a lot of data breaches, personal information loses and a lot more.

## *RECOMMENDATIONS, IMPLICATIONS AND CONTRIBUTIONS*

According to the study's findings and results, the researcher comes up with some recommendations, suggestions and implications that hoteliers can implement to make their computer system and networks more secure and to prevent any type of data breach and cyber attacks as much as possible:

There are variety of tools and techniques available to scan the vulnerability of the computer networks, measure and protect data from cyber attacks (as some of them are mentioned in this research), that any hotel based on its affordability, size and other features of the hotel, can implement and use in order to take a better care of the data and personal information of the guests. Also each hotel should have a contract with an IT company or a dedicated IT manager whom the hotel trusts so that on a regular bases all the computer networks can be checked for any possible problem that makes the computer networks vulnerable.

In addition to that, hotels should determine some internal regulations and policies for the hotel's employees regarding cyber security and computer network usage as well as having cyber security training program for those employees whose job is working with computers, handling the emails and social media. Moreover, instead of having the same password for all the guests in the hotel or not having any password to join the hotel Wi-Fi, hotels can set up a wired internet system, so that those guests who has concern about their security be able to use the wired system while they are in their room or having an individual password for each guest or each room in order to protect the computer networks from any cyber attack that might be caused by reusable passwords. Furthermore, hotels must have a very secure and certified website that leverage extended validation or at least domain validation, so that the guests be able to book the rooms or anything else online without having any concern of being hacked or abused. And finally it is better for any hotel to have cyber insurance so that if the hotel experienced any data breach or cyber attack, the insurance covers its loss and liabilities.

Although this study has many more contribution, the main contributions of this study are increasing the public knowledge and awareness regarding the importance of cyber security in hotels and hospitality industry as well as the importance of having a strong password for the online accounts and changing them more often. It also teaches people to be careful about the security of the website they want to pay online while booking a hotel room. In addition, this study provides a thorough discussion of important terminologies and concepts of cyber security in general and hospitality industry in particular. And finally this study helps to increases the knowledge of the readers about the different tools and techniques of cyber security as well as their usage.

## *LIMITATIONS AND FUTURE STUDIES*

Considering that no research can be done without limitations, this study also has several limitations. According to Phillips (1981), the best participants for interviews in research are those who have enough information and knowledge about the topic and issue that the researcher is seeking for. Based on that, unfortunately the researcher of this study was not able to find exactly those participants, who are expert in this area since three out of five hotels, which were targeting for this study, didn't have IT manager in their hotel and the interview was done with the assistant of general managers in those three hotels. In addition, assistant of general managers did not have much information and knowledge about the issue.

On the other hand, the biggest limitation that this study faced was that in all of the five hotels that the interviews were done, due to hotels' policy and the other involved companies' policies (e.g., the policies of the security companies involved), the participants were not allowed to reveal information about the hotel's security. The hotel employee interviewees' mentioned that it was against their hotel's policy to talk about the hotel's network security as well as their experiences regarding the security threats and challenges faced by the hotel. Therefore, the researcher was not able to get the answer of all the interview questions.

Another limitation of this study is that the sample of the study was very general and did not target any specific features in choosing the considered five hotels, such as property features, number of stars, services included, etc.

Furthermore, this study has only focused on network security and customers' data and information security and not specifically on PCI DSS, payment security methods or anything regarding credit cards and debit cards fraud. However, the project did briefly define the terminologies regarding other security aspects. In particular, the project described some details of PCI DSS and its purpose in businesses, organizations and specifically hotels. Furthermore, the project discussed the determination of the compliance level of PCI DSS in a business or an organization. Thus, the future studies can focus more on how the PCI DSS works in hotels and base on the size of the hotels and the other features of them, which level of compliance they need to have.

Also, in this study, the researcher shortly introduced tokenization, which can be an interesting topic for future studies to see if hotels really know about this issue and take advantage of it to secure their network system or not and how it can be a help for them to prevent data breach and cyber attacks.

Since network security and cyber crimes are kind of new topics and hotels are facing more and more challenges as well as getting to know the importance of this issue gradually, a lot of research can be done in this area. For example one of the targets of the research can be the insurance companies that cover liabilities regarding cyber crimes in hotels and other organizations. In addition, similar research can be carried out in other industries, such as banking and airlines.